\documentclass[aps,pre,twocolumn,showpacs]{revtex4}
\usepackage{epsfig}
\usepackage{times}
\begin{document}

\title{Phase diagrams for the spatial public goods game with pool-punishment}

\author{Attila Szolnoki,$^1$ Gy\"orgy Szab\'o,$^1$ and Matja{\v z} Perc,$^2$}
\affiliation{$^1$ Research Institute for Technical Physics and Materials Science, P.O. Box 49, H-1525 Budapest, Hungary\\
$^2$ Faculty of Natural Sciences and Mathematics, University of Maribor, Koro{\v s}ka cesta 160, SI-2000 Maribor, Slovenia}

\begin{abstract}
The efficiency of institutionalized punishment is studied by evaluating the stationary states in the spatial public goods game comprising unconditional defectors, cooperators, and cooperating pool-punishers as the three competing strategies. Fine and cost of pool-punishment are considered as the two main parameters determining the stationary distributions of strategies on the square lattice. Each player collects its payoff from five five-person public goods games, and the evolution of strategies is subsequently governed by imitation based on pairwise comparisons at a low level of noise. The impact of pool-punishment on the evolution of cooperation in structured populations is significantly different from that reported previously for peer-punishment. Representative phase diagrams reveal remarkably rich behavior, depending also on the value of the synergy factor that characterizes the efficiency of investments payed into the common pool. Besides traditional single and two-strategy stationary states, a rock-paper-scissors type cyclic dominance can emerge in strikingly different ways.
\end{abstract}

\maketitle

\section{Introduction}
\label{intro}

The importance of punishment for the maintenance of cooperative behavior in human societies can be quantified by studying spatial public goods games (PGG) with players forming overlapping groups. For the simple two-strategy case players (within all the groups) decide simultaneously whether they wish to contribute to the common pool (cooperate) or not (defect). Subsequently, the multiplied total investment is divided equally among all the group members irrespectively of their initial decision. In this situation the rational (selfish) players should decline to contribute if the investment costs exceed the return of the game \cite{nowak_06, sigmund_10}. As a result, selfish players fail to benefit from mutual cooperation and the society evolves towards the "tragedy of the commons" \cite{hardin_g_s68}. Human experiments and mathematical models alike have shown, however, that cooperative behavior can be promoted by punishing defectors for a wide class of social dilemmas, including the prisoner's dilemma game. In fact, it can be stated that some elements of punishment can be recognized within all the relevant mechanisms \cite{nowak_s06} supporting cooperation among selfish individuals \cite{guan_pre07,wakano_pnas09, liu_rr_pa10, zhang_jl_pa10}.

Traditionally, the sanctions foreseen by punishment are considered to be costly. While those that are punished bear a fine, the punishers must bear the cost of punishment. Both fine and cost may substantially reduce the overall income of the corresponding players. There are, however, different ways of how the income reduction is executed that depend on the governing evolutionary rules, the set of strategies, as well as on the network structure and group formation, among others. Many aspects of punishment were already investigated by experiments \cite{clutton_brock_n95, fehr_n02, fehr_n03, semmann_n03, de-quervain_s04, fowler_pnas05, henrich_s06, sasaki_prsb07, egas_prsb08}, as well as by means of mathematical models with three \cite{hauert_s02, szabo_prl02, bowles_tpb04, brandt_pnas05}, four \cite{sigmund_pnas01, ohtsuki_n09}, and even more strategies \cite{henrich_jtb01, dreber_n08}.

Here we study the effects of pool-punishment in the spatial PGG and contrast the results with those reported previously for peer-punishment \cite{helbing_njp10, helbing_ploscb10, helbing_pre10c}. Pool-punishment is synonymous to institutionalized punishment, where the contributions of punishers are meant to cover the costs of institutions like the police or other elements of the justice system independently of their necessity or efficiency \cite{sigmund_n10}. On the contrary, by peer-punishment \cite{fehr_n02, gardner_a_an04} the punishers pay the cost of punishment only if it is necessary, \textit{i.e.} when the defectors are identified and sanctioned. In the absence of defectors the income of peer-punishers is therefore equivalent to that of pure cooperators, who refuse to bear the cost of punishment and are thus frequently referred to as the ``second order free-riders". On the other hand, because of their permanent contributions to the punishment pool the income of pool-punishers is always smaller than that of cooperators. A preceding study on pool-punishment in well-mixed populations \cite{sigmund_n10} concluded that pool-punishers can prevail over peer-punishers only if the second-order free-riders are punished as well. We will show that in structured populations self-organizing spatiotemporal structures can maintain pool-punishment viable without such an assumption. Indeed, the phase diagrams for three representative values of the multiplication parameter at a low level of noise indicate surprisingly rich behavior depending on the punishment fine and cost.

\section{Spatial public goods game with pool-punishment}
\label{model}

The PGG is staged on a square lattice with periodic boundary conditions. The players are arranged into overlapping five-person ($G=5$) groups in a way such that the focal players are surrounded by their four nearest neighbors each. Accordingly, each individual belongs to $G=5$ different groups. All the players thus play five five-person PGGs by following the same strategy in every group they are affiliated with. Initially each player on site $x$ is designated either as a pool-punisher ($s_x = O$), cooperator ($s_x = C$), or defector ($s_x = D$) with equal probability. Using standard parametrization, the two cooperating strategies $O$ and $C$ contribute a fixed amount (here considered being equal to $1$ without loss of generality) to the public good while defectors contribute nothing. The sum of all contributions in each group is multiplied by the factor $1<r<G$, reflecting the synergetic effects of cooperation, and the resulting amount is then equally divided among all the group members irrespective of their strategies.

Pool-punishment requires precursive allocation of resources and therefore each punisher contributes an amount $\gamma$ to the punishment pool irrespective of the strategies in its neighborhood. Defectors, on the other hand, must bear the punishment fine $\beta$, but only if there is at least one pool-punisher present in the group. Denoting the number of cooperators ($C$), pool-punishers ($O$) and defectors ($D$) in a given group $g$ by $N_{C}^g$, $N_{O}^g$ and $N_{D}^g$, respectively, the payoffs
\begin{eqnarray}
P_{C}^g&=&r(N_{C}^g+N_{O}^g)/G - 1, \nonumber \\
P_{O}^g&=&P_{C}^g - \gamma , \\
P_{D}^g&=&r(N_{C}^g+N_{O}^g)/G-\beta f(N_{O}^g)  \nonumber
\label{eq:payoffs}
\end{eqnarray}
are obtained by each player $x$ depending on its strategy $s_x$, where the step-like function $f(Z)$ is $1$ if $Z>0$ and $0$ otherwise.

By Monte Carlo (MC) simulations the system is started from a random initial strategy distribution, and its evolution is subsequently controlled by the more realistic random sequential strategy updates \cite{roca_plr09}. During these elementary processes a randomly selected player $x$ plays the public goods game with its interaction partners as a member of all the $g=1, \ldots, G$ groups, whereby its overall payoff is thus $P_{s_x} = \sum_g P_{s_x}^g$. Next, player $x$ chooses one of its four nearest neighbors at random, and the chosen co-player $y$ also acquires its payoff $P_{s_y}$ in the same way. Finally, player $x$ imitates the strategy of player $y$ with a probability $w(s_x \to s_y)=1/\{1+\exp[(P_{s_x}-P_{s_y})/K]\}$, where $K$ quantifies the uncertainty in strategy adoptions \cite{szabo_pre98}. Without loss of generality we set $K=0.5$, thereby allowing also direct comparisons with previous results obtained for the same level of noise \cite{helbing_ploscb10, helbing_njp10}. Each Monte Carlo step (MCS) (interpreted as a unit of time) gives a chance for the players to adopt a strategy from one of their neighbors once on average. The average frequencies of pool-punishers ($\rho_{O}$), cooperators ($\rho_{C}$) and defectors ($\rho_{D}$) on the square lattice are determined in the stationary state after a sufficiently long relaxation time $t_r$. Depending on the actual conditions (proximity to phase transition points and the typical size of emerging spatial patterns) the linear system size was varied from $L=200$ to $5000$, and both the relaxation ($t_r$) and the sampling ($t_s$) time were varied from $t_r \simeq t_s = 10^4$ to $10^7$ MCS to ensure that the statistical error is comparable with the line thickness in the plots.

The first study of the spatial two-strategy ($D$ and $C$) evolutionary prisoner's dilemma games (PDGs) indicated that the survival of cooperators is supported if they form compact clusters \cite{nowak_n92b}. Similar phenomena were subsequently reported for spatial evolutionary PGGs \cite{szabo_prl02, roca_pre09}. On the contrary, the survival of defectors is enhanced if they are distributed sparsely. Quantitative analyses have revealed that cooperators and defectors coexist in the stationary state if $r_{c1} < r < r_{c2}$ (henceforth this state will be denoted as DC), where the two threshold values depend on the connectivity structure (including the group size $G$) and the noise level. Below (above) the borders of the coexistence phase only defectors (cooperators) remain alive, while within the DC region $\rho_C$ increases monotonously from $0$ to $1$. It turned out, furthermore, that for the spatial PGG the extension of the coexistence region ($r_{c2}-r_{c1}$) remains finite in the zero noise limit for all the previously studied connectivity structures \cite{szolnoki_pre09c}. This is in sharp contrast with the results obtained for spatial PDGs, where $r_{c2}-r_{c1} \to 0$ in the $K \to 0$ limit for several connectivity structures (\textit{e.g.} on the square lattice) \cite{szabo_pre05, vukov_pre06}. Consequently, in our simulations the noise level $K=0.5$ yields a typical low noise behavior with a sufficiently fast relaxation towards the final stationary state.

\begin{figure*}[ht!]
\centerline{\epsfig{file=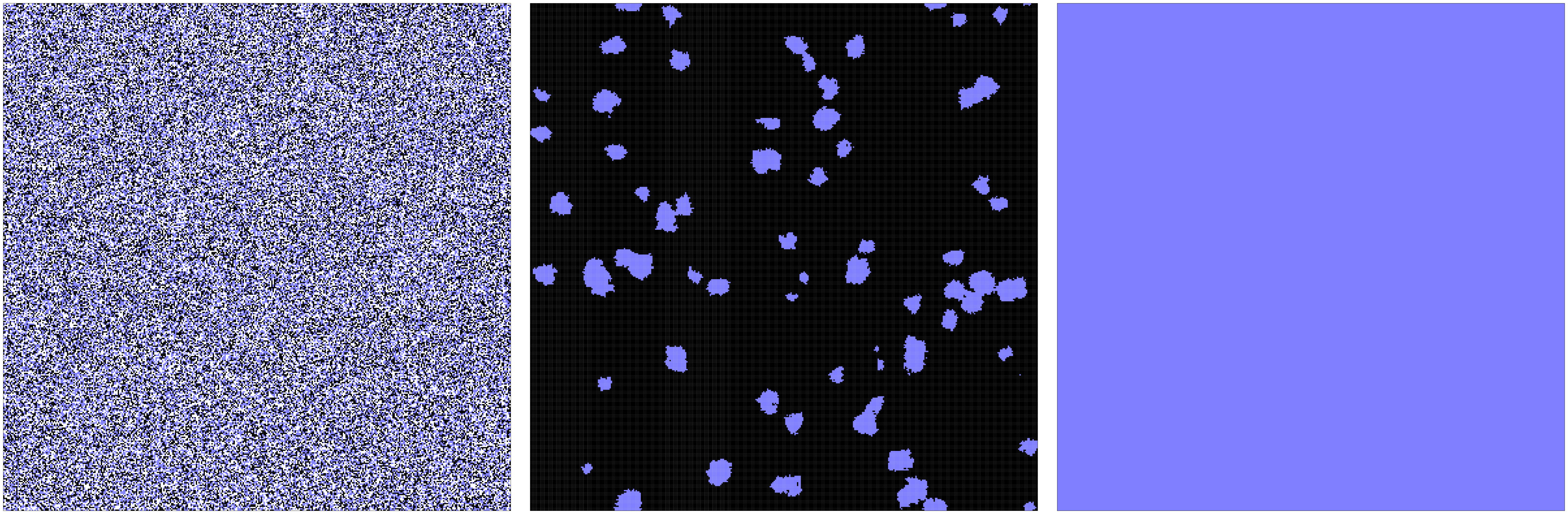,width=15.0cm}}

\hspace{1cm}

\centerline{\epsfig{file=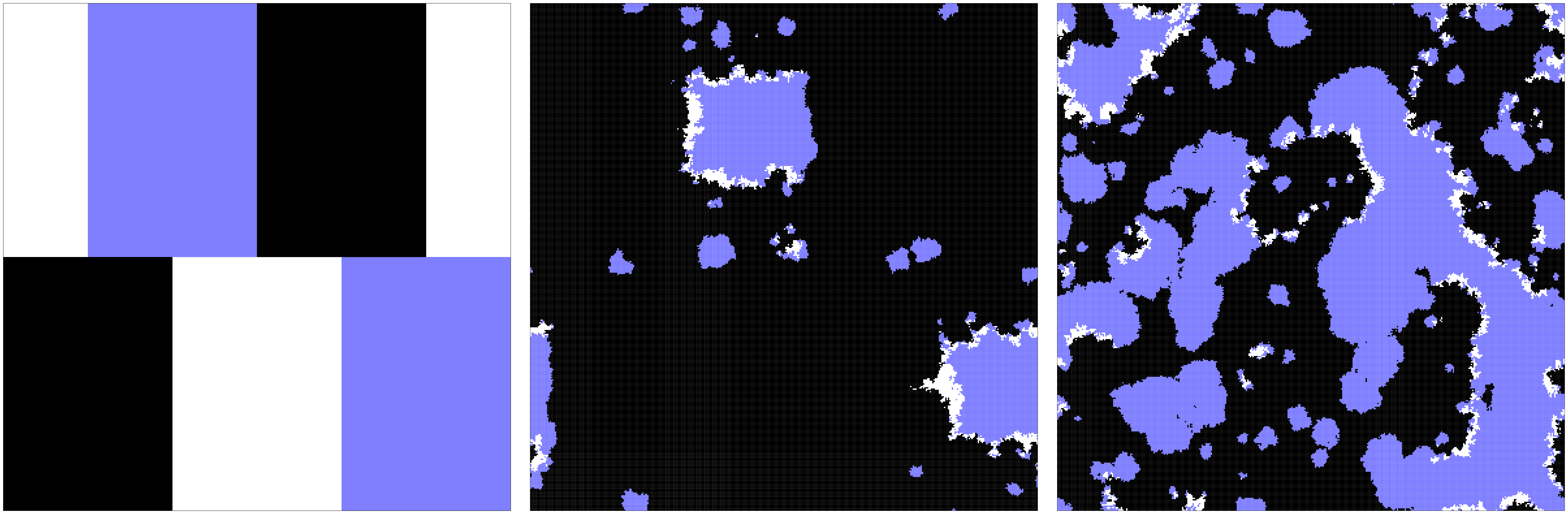,width=15.0cm}}

\hspace{1cm}

\centerline{\epsfig{file=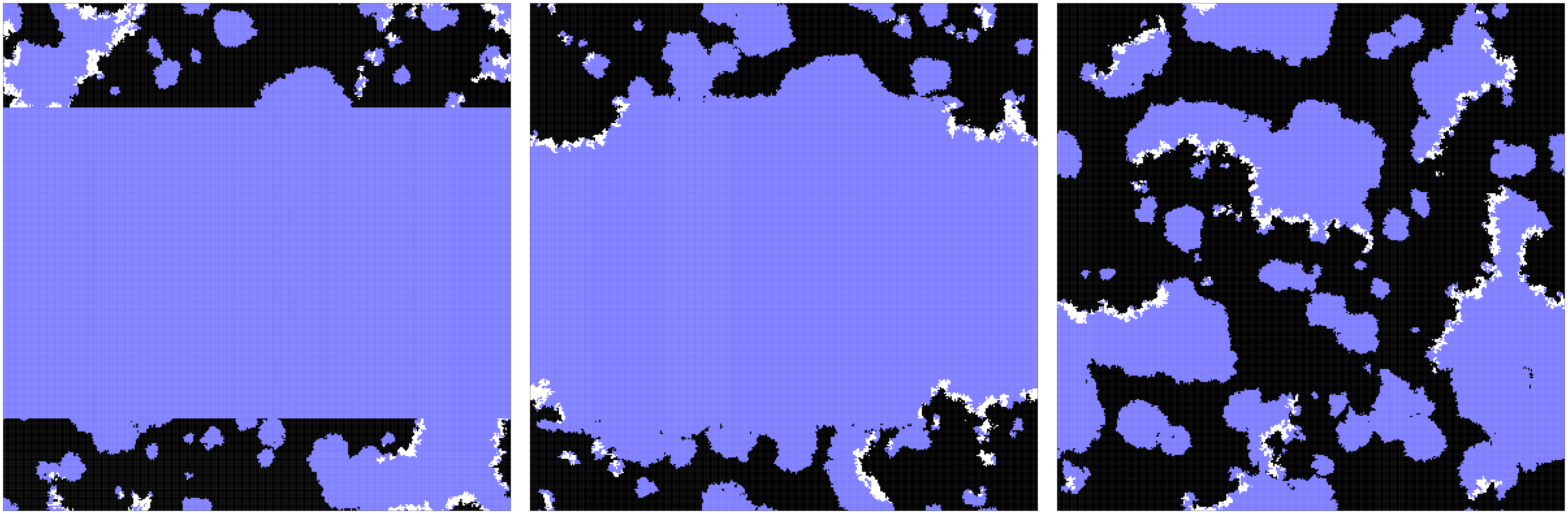,width=15.0cm}}

\caption{(Color online) Evolution of strategy distribution for three different initial states. Upper row shows the evolution from random initial state while middle row shows the evolution from a prepared state. The bottom row demonstrates the time evolution when the final states of upper evolutions meet. We used identical parameters for all cases, namely $L=390$, $r=2.0$, $\gamma=0.1$, $\beta=0.79$ and $K=0.5$. Black, white, and blue (grey in black\&white print) denotes players with defector, pure cooperator, and pool-punisher strategy, respectively.}
\label{multi}
\end{figure*}

Evidently, if only pure cooperators and pool-punishers are initially present in the system, then all the pool-punishers will eventually die out because of their lower payoff. At the same time, analogous with the coexistence of the $D$ and $C$ strategies, the $D$ and $O$ players can also coexist in the so-called DO phase that is bounded to a synergy factor region $r_{c1}^{\prime} < r < r_{c2}^{\prime}$, where the two threshold values are affected not only by the connectivity structure and the noise level, but also by the values of the punishment fine $\beta$ and cost $\gamma$.

We emphasize that the homogeneous one-strategy solutions, denoted henceforth as C, D and O phases, are absorbing states because the applied dynamical rule leaves these states unchanged once the system arrives there. Due to the analogy between the presently applied imitation rule and the spreading of infections, simplified by the contact process \cite{liggett_85}, it is expected that upon varying one of the parameters the above-mentioned continuous phase transitions from a two-strategy state to one of the homogeneous phases will belong to the directed percolation universality class. Up to now this was confirmed only for the spatial PDGs \cite{szabo_pre98, chiappin_pre99} (for further references see \cite{szabo_pr07}). In the following sections the power law behavior of the extinction process is verified only for a few cases because of the huge computational capacity that is required for this. Similar critical transitions can also be observed when a three-strategy state transforms into a two-strategy state by varying a control parameter. Such behavior was already observed previously in a spatial evolutionary PGG with volunteering \cite{szabo_prl02}.

In the majority of spatial systems the three-strategy states are maintained by cyclic dominance among the three strategies. Examples include the PGG \cite{hauert_s02, szabo_prl02} and the PDG with voluntary participation \cite{szabo_pre02b}, as well as other three-strategy (\textit{e.g.} cooperation, defection and tit-for-tat) variants of the PDG \cite{hutson_prsb95, reichenbach_jtb08, zhang_gy_pre09, szolnoki_pre10b}. In the spatial PGG with pool-punishment, the cooperators can invade the territory of punishers, the punishers can occupy the sites of neighboring defectors, while defectors may outperform cooperators within a wide range of parameters. We find that this rock-paper-scissors type cyclic dominance yields a self-organizing pattern, which we will henceforth denote by (D+C+O)$_c$. We emphasize that an analogous three-strategy phase governed by cyclic dominance cannot be observed if peer-punishment is considered, because there, in the spatial mixture of cooperators and peer-punishers, both types of players receive the same payoff, and furthermore, due to the random imitation the evolution of the system becomes equivalent to that of the voter model \cite{liggett_85}. Interestingly though, the extinction of free-rider pure cooperators can be catalyzed efficiently by adding defectors via rare random mutations \cite{helbing_pre10c}. The description and notation of additional phases including the governing phase transitions will be given below at the place of their occurrence.

As a general comment for the simulation difficulties of spatial system of cyclic dominant species, we must highlight the potential problem originated form small system size. If the system size is not large enough then the simulations can result one- and/or two-strategy solutions that are unstable against the introduction of a group of mutants. For example, the homogeneous C or O phases can be invaded completely by the offspring of a single defector inserted into the system at sufficiently low values of $r$. On the contrary, the D phase can be fully occupied by a single group of pool-punishers (or cooperators) if initially they form a sufficiently large compact cluster (\textit{e.g.} a rectangular box). In such cases the competition between two homogeneous phases can be characterized by the average velocity of the invasion fronts separating the two spatial solutions characterized by a proper composition and spatiotemporal structure. Generally, the same method can also be used to determine the winner between any two possible spatial solutions.

For the considered imitation rule a system with three (or more) strategies has a large number of possible solutions because all the solutions of each subsystem (comprising only a subset of all the original strategies) are also solutions of the whole system \cite{szabo_pr07}. In such situations the most stable solution can be deduced by performing a systematic check of stability (direction of invasion) between all the possible pairs of subsystem solutions that are separated by an interface in the spatial system. Fortunately, this analysis can be performed simultaneously if we choose a suitable patchy structure of subsystem solutions where all the possible interfaces are present. The whole grid is then divided into several large rectangular boxes with different initial strategy distributions (containing one, two or three strategies), and the strategy adoptions across the interfaces are initially forbidden for a sufficiently long initialization period. By using this approach one can avoid the difficulties associated either with the fast transients from a random initial state or with the different time scales that characterize the formation of possible subsystem solutions. On the contrary, it is easy to see that a random initial state may not necessarily offer equal chances for every solution to emerge. Evidently, if the system size is large enough then these solutions can form locally and the most stable one can subsequently invade the whole system. At small system sizes, however, only those solutions can evolve whose characteristic formation times are short enough.

To illustrate the possible problem of random initial states when using small system sizes, we compare the time evolution of strategy distributions for different initial states in Fig.~\ref{multi}. For appropriate comparisons, naturally, we have used identical model parameters for all cases, namely synergy factor $r=2.0$, the cost of punishment $\gamma=0.1$, and the fine $\beta=0.79$ at system size $L=390$. The upper three snapshots demonstrate that the system arrives to the O phase when the strategies are initially distributed randomly (snapshots are given at $t=0, 100$, and $1000$ MCS). The middle panel demonstrates what happens if the initial state (left side) contains all the possible interfaces and vertices of homogeneous domains of the three strategies (snapshots are given at $t=0, 200$, and $3750$ MCS). Here, the right plot illustrates the (D+C+O)$_c$ stationary state in which all the three strategies are present due to cyclic dominance. The bottom panel shows the competition of O and (D+C+O)$_c$ states where the latter is the winner (snapshots are given at $t=0, 100$, and $1600$ MCS). As we have already mentioned, the most stable (D+C+O)$_c$ state can also emerge and spread from a random initial state, but only if the system size is large enough [in case of random initial conditions the system size should exceed $L=1500$ for these $(r, \gamma, \beta)$ parameter values to obtain a reliable solution].

\section{Results of Monte Carlo simulations}
\label{results}

Systematic MC simulations are performed to reveal phase diagrams for three representative values of the multiplicative factor $r$. In each case we have determined the stationary frequencies of strategies when varying the fine $\beta$ for many fixed values of cost $\gamma$ ($\beta, \gamma > 0$). The transition points and the type of phase transitions are identified from MC data collected with a sufficiently high accuracy (and frequency) in the close vicinity of the transition points. Finally, the phase boundaries, separating different stable solutions, are plotted in the full fine-cost phase diagrams.

The three values of $r$ give rise to fundamentally different behavior when increasing the value of fine. In the first two cases ($r=2$ and $3.5$) the cooperators die out in the absence of pool-punishment \cite{szolnoki_pre09c}. The third value of the synergy factor ($r=3.8$) is chosen to illustrate the impact of punishment when $C$ and $D$ coexist in the absence of $O$. Evidently, the consideration of punishment becomes futile when the cooperators beat defectors in the absence of punishers ($r>r_{c2}$). When $r$ increases towards $r_{c2}$ the effect of punishment decreases with $\rho_D$ (within the DC phase). The obtained quantitative results are discussed in detail in the following three subsections.

\subsection{Results for the synergy factor $r=2.0$}
\label{results20}

\begin{figure}
\centerline{\epsfig{file=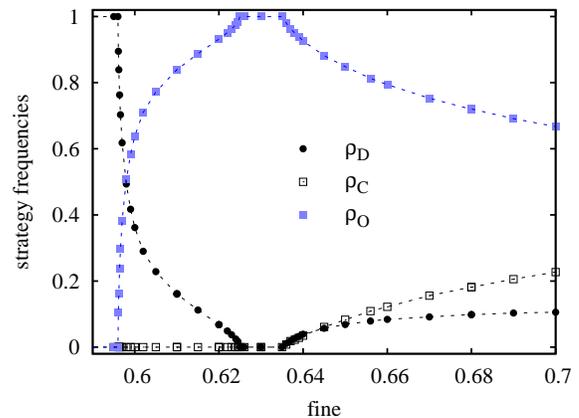,width=7.6cm}}
\caption{(Color online) Strategy frequencies {\it vs}. fine $\beta$ for the punishment cost $\gamma=0.01$ at $r=2.0$ and $K=0.5$.}
\label{cross_r2a}
\end{figure}

First, we illustrate the variation of strategy frequencies and also the phase transitions obtained by means of MC simulations as a function of fine for a low value of cost. Figure \ref{cross_r2a} shows consecutive transitions from the pure D phase to the final (D+C+O)$_c$ phase described above.

When increasing the fine $\beta$ at a low value of cost ($\gamma=0.01$) one can observe three continuous phase transitions. First, the homogeneous defector state (D) transforms into the coexistence of defectors and pool-punishers (DO). In this phase, pool-punishers form compact clusters to survive in the sea of defectors. This mechanism is identical to the previously identified network reciprocity that enables pure cooperators to coexist with defectors \cite{nowak_n92b}. Cooperators who refuse to bear the cost of punishment, however, are unable to survive due to the low value of $r$. Within the DO phase the frequency of pool-punishers increases continuously until the homogeneous O phase is reached. Surprisingly, further increasing $\beta$ induces an additional phase transformation from the O phase into the (D+C+O)$_c$ phase, where the self-organizing pattern is maintained by cyclic dominance and the nonzero frequencies for all strategies remain valid in the large fine limit.

\begin{figure}
\centerline{\epsfig{file=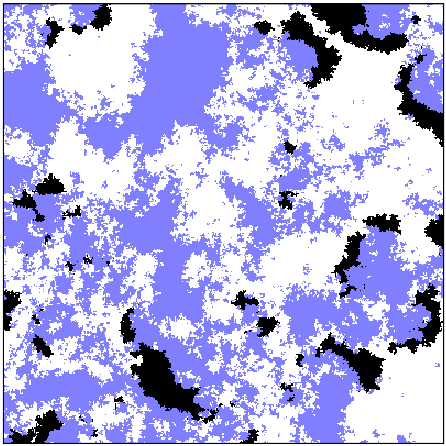,width=6cm}}
\caption{(Color online) Typical distribution of strategies on a $400 \times 400$ portion of a larger square lattice for $r=2.0$, $\beta=0.01$ and $\gamma=1.0$. The color code is the same as used in Fig.\ref{multi} and for symbols in Fig.~\ref{cross_r2a}.}
\label{snapsh1}
\end{figure}

Within the (D+C+O)$_c$ phase $\rho_O$ decreases monotonously with $\beta$ in agreement with the anomalous behavior referred frequently as the ``survival of the weakest'' \cite{tainaka_pla93}. In the present case, the increase of fine reduces the income of the punished defectors, which allows pure cooperators to survive. The latter strategy behaves as the ``predator'' of pool-punishers, resulting in the decay of $\rho_O$ despite of the increasing fine. The same cyclic dominance mediated complex interaction is able to increase $\rho_O$ when $\gamma$ is increased (in this case the less effective punishment does not allow $C$ players, who are the ``prey'' of $D$, to survive). Similar effects were already reported in several three-strategy models, including the simpler spatial rock-paper-scissors game, and the main features were justified by mean-field approximations and pair-approximations (for a brief survey see the review \cite{szabo_pr07} and further references therein). The robustness of this behavior can be demonstrated effectively by a snapshot (see Fig.~\ref{snapsh1}), illustrating significantly different interfaces between the coexisting phases.

At such a low punishment cost the cooperators can invade the sites of pool-punishers, albeit very slowly and only within the territories they have in common. It is emphasized that within these two-strategy territories in the $\gamma \to 0$ limit the strategy evolution reproduces the behavior of the voter model with equivalent strategies exhibiting rough interfaces and extremely slow coarsening \cite{dornic_prl01}. For low but finite values of $\gamma$ the two-strategy system evolves slowly towards the homogeneous C state while the interfaces remain irregular as demonstrated in Fig.~\ref{snapsh1}. Notice that the interfaces separating the domains of defectors from cooperators or defectors from pool-punisher are less irregular, thus signaling the more obvious dominance between these strategy pairs.

\begin{figure}
\centerline{\epsfig{file=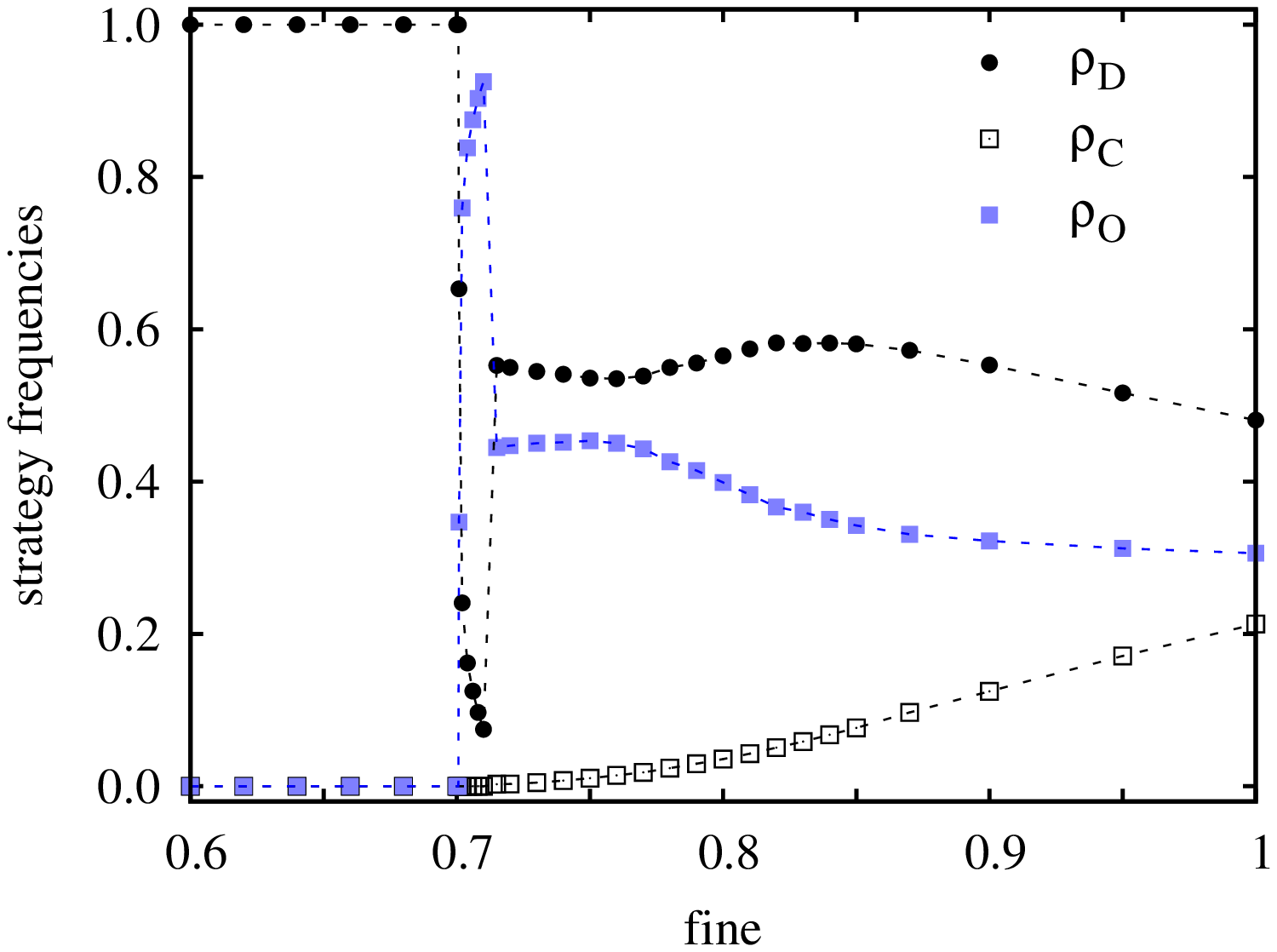,width=7.5cm}}
\centerline{\epsfig{file=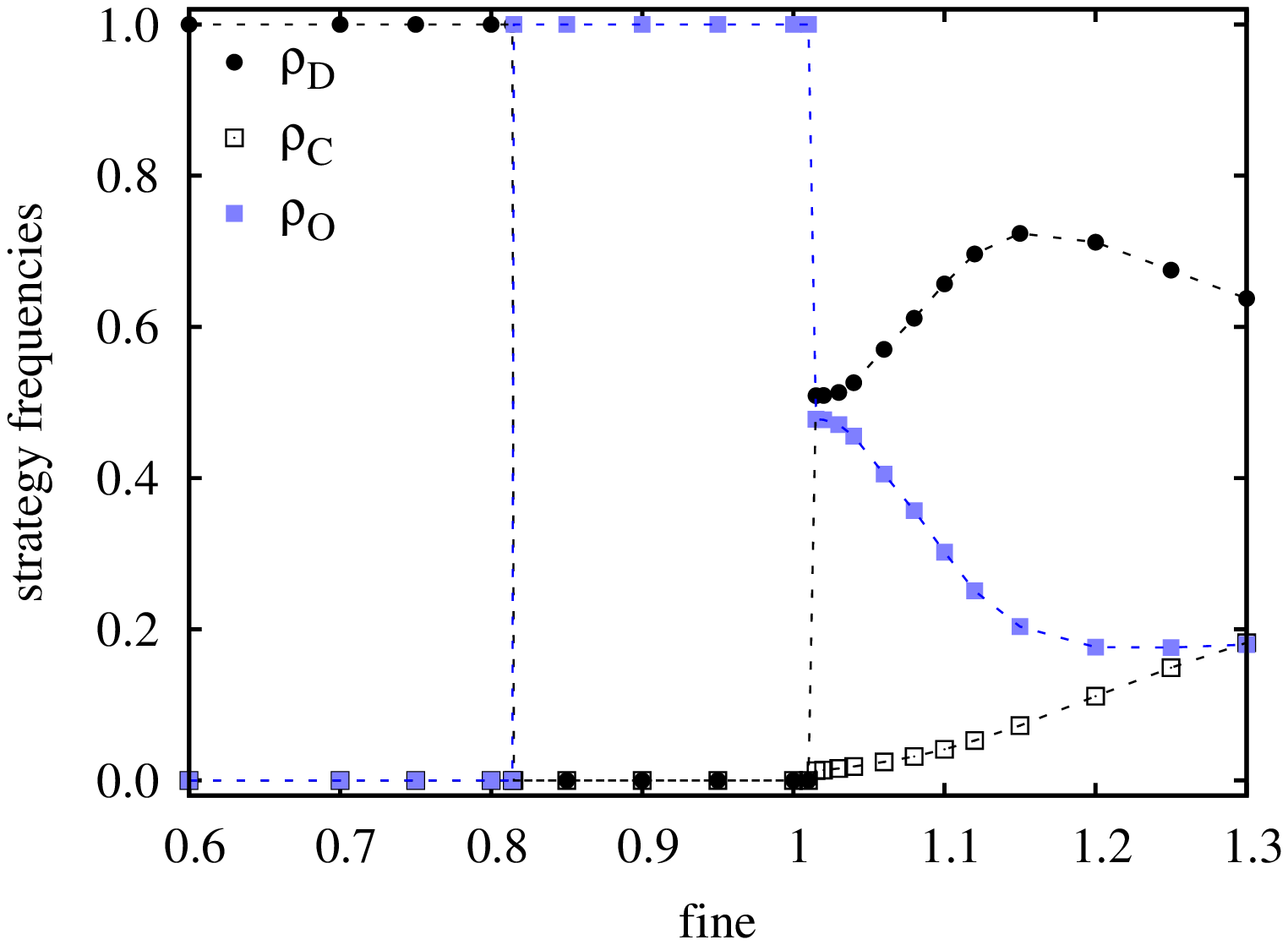,width=7.5cm}}
\caption{(Color online) Strategy frequencies {\it vs}. fine $\beta$ for the punishment cost $\gamma=0.1$ (top) and $\gamma=0.2$ (bottom) at $r=2.0$ and $K=0.5$.}
\label{cross_r2b}
\end{figure}

The increase of the punishment cost $\gamma$ reduces the net income of pool-punishers, consequently yielding fundamentally different variations in the strategy frequencies upon increasing of the fine $\beta$, as demonstrated in Figs.~\ref{cross_r2b}. The upper plot illustrates the disappearance of the pure O phase if $\gamma=0.1$. The extension (along $\beta$) of the DO phase decreases linearly with $\gamma$ and vanishes at $\gamma = 0.212$. At the same time, the homogeneous O phase can also be observed between the phases D and (D+C+O)$_c$, but only if the cost exceeds a threshold value [here $\gamma>\gamma_{th1}(r=2)=0.113$] that depends also on the multiplication factor $r$.

Notice that for both values of the punishment cost $\gamma$ the (D+C+DO)$_c$ phase occurs via a first-order (discontinuous) phase transition when $\beta$ increases, as can be inferred from the two panels of Fig.~\ref{cross_r2b}. Furthermore, the transition from D $\to$ O also becomes discontinuous in the absence of the DO phase.

\begin{figure}
\centerline{\epsfig{file=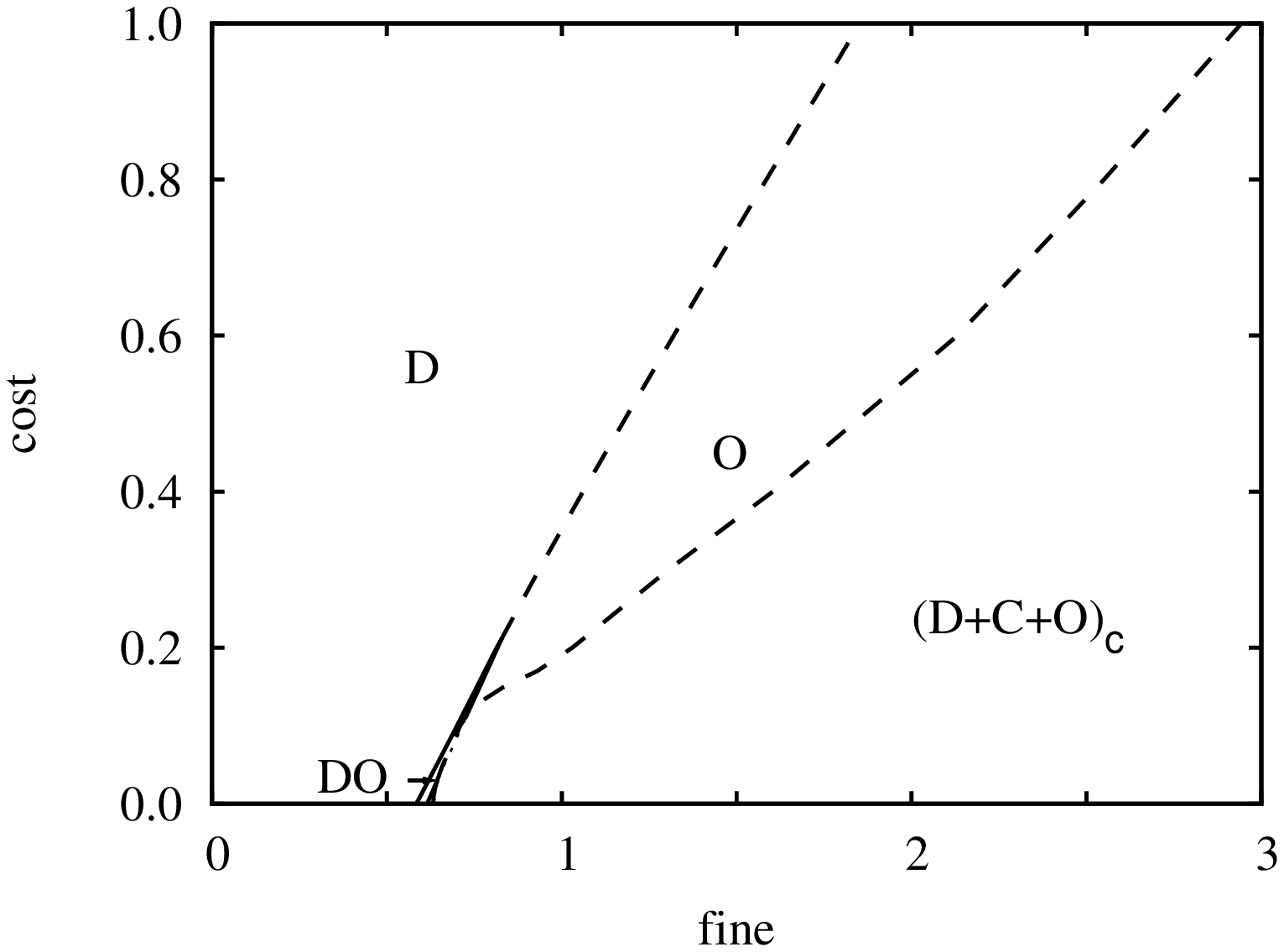,width=7.5cm}}
\centerline{\epsfig{file=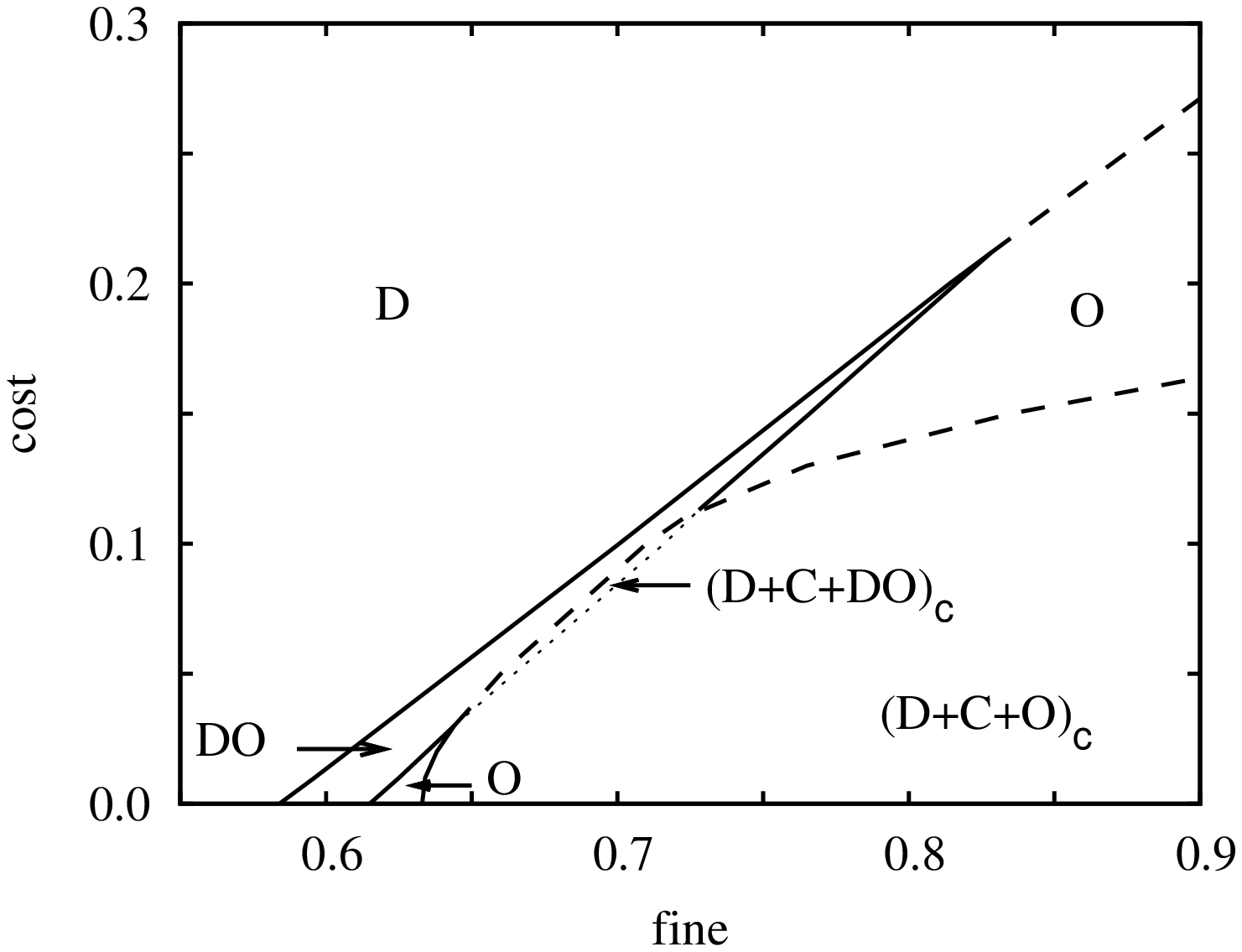,width=7.7cm}}
\caption{Full fine-cost phase diagram for $r=2.0$ and $K=0.5$. Bottom panel features the enlargement of the small-cost area. Solid (dashed) lines indicate continuous (discontinuous) phase transitions. The dotted line represents the analytic continuation of the phase boundary separating the pure D and O phases in the absence of cooperators ($C$).}
\label{phd_r2}
\end{figure}

The above numerical investigations were repeated for many other values of $\gamma$, and the results are summarized in the full fine-cost phase diagram presented in Fig.~\ref{phd_r2}, where the lower plot magnifies the most complex (small-cost) region. The lower (magnified) phase diagram refers to an additional new phase [(D+C+DO)$_c$ marked with an arrow], which we will, however, address in the following section because it has more obvious consequences at higher values of $r$.

In general, the presented fine-cost phase diagram shows clearly that in this low-$r$ region only defectors remain alive if the fine does not exceeds a threshold value that increases approximately linearly with the cost of punishment. More precisely, in the low noise limit the phase boundary separating the D and O phases approaches the straight line with a slope of $4/5$, and this boundary moves left if $r$ is increased.

\subsection{Results for the synergy factor $r=3.5$}
\label{results35}

\begin{figure}
\centerline{\epsfig{file=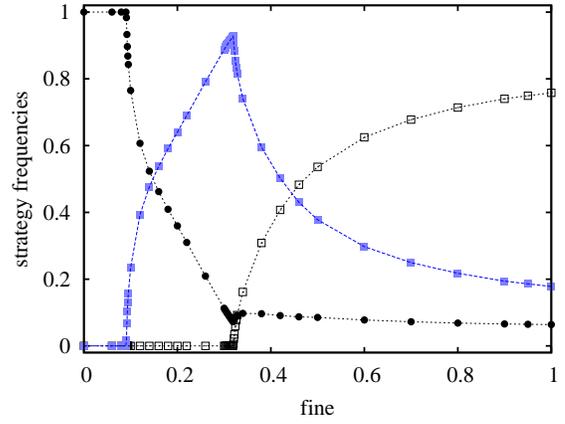,width=7.5cm}}
\caption{(Color online) Strategy frequencies {\it vs}. fine $\beta$ for the punishment cost $\gamma=0.01$ at $r=3.5$ and $K=0.5$.}
\label{csr35c01}
\end{figure}

Here for the low cost limit the system behavior is similar to the one described in the previous section. The relevant difference is the absence of the O phase in the series of transitions upon increasing the fine $\beta$, as demonstrated in Fig.~\ref{csr35c01}. Notice that both transitions are continuous. The quantitative analysis supports the conjecture \cite{janssen_zpb81} that the continuous extinction of either the pool-punishers or the cooperators belongs to the directed percolation universality class.

\begin{figure}
\centerline{\epsfig{file=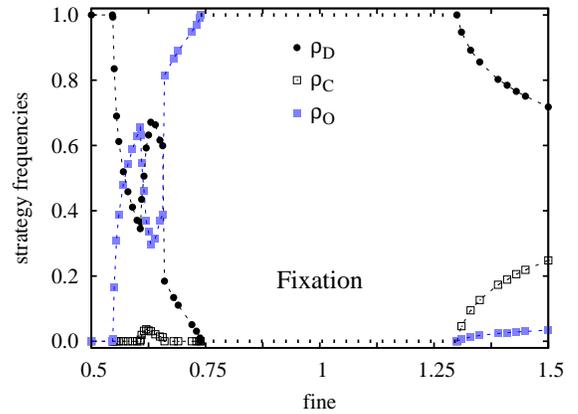,width=7.5cm}}
\caption{(Color online) Strategy frequencies {\it vs}. fine $\beta$ for the punishment cost $\gamma=0.4$ at $r=3.5$ and $K=0.5$.}
\label{csr35c4}
\end{figure}

Figure~\ref{csr35c4} shows relevant differences in the fine-dependence of the strategy frequencies for higher values of $\gamma$. In this case our simulations indicate four intermediate phases between the phases D and (D+C+O)$_c$ if the fine is increased at a fixed cost and noise level. The five critical points for the consecutive transitions will be denoted as $\beta_{c1} < \ldots < \beta_{c5}$. For example, the first transition at $\beta=\beta_{c1}$ refers to a continuous transition from the phase D to DO in agreement with the cases discussed above.

\begin{figure}
\centerline{\epsfig{file=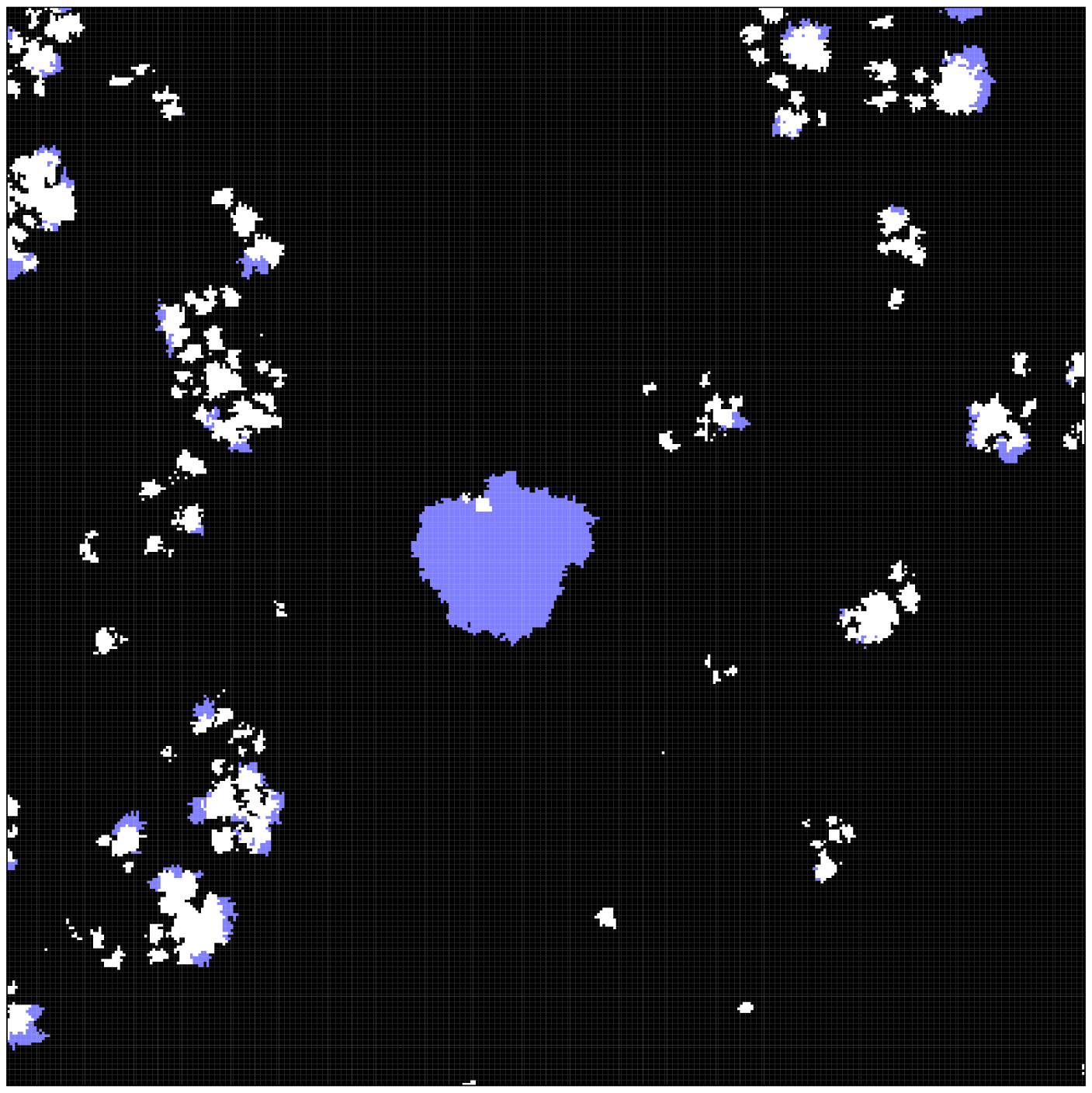,width=6cm}}
\caption{(Color online) Typical distribution of strategies on a $400 \times 400$ portion of a larger square lattice for $r=3.5$, $\beta=0.2$ and $\gamma=0.78$. The color code is the same as used in Fig.\ref{multi} and for symbols in Figs.~\ref{csr35c01} and \ref{csr35c4}.}
\label{snapsh2}
\end{figure}

We first emphasize a striking novel feature in the fine-dependence of the strategy frequencies within the (D+C+O)$_c$ phase [$\beta > \beta_{c5}(r=0.4)=1.30(1)$]. In this case $\rho_D \to 1$ (and evidently, both $\rho_C$ and $\rho_O$ converge to zero) if $\beta$ approaches $\beta_{c5}$ from higher values in contrary to all the previous cases demonstrated in Figs.~\ref{cross_r2a}, \ref{cross_r2b} and \ref{csr35c01}. This behavior is accompanied with a drastic change in the governing spatial patterns, as demonstrated in Fig.~\ref{snapsh2}.

Namely, in the close vicinity of the transition point smaller or larger islands of cooperators and/or pool-punishers are dispersed in the see of defectors. Due to the cyclic dominance the islands of pool-punisher are blowing up while the cooperator islands are shrinking and disappear in most of the cases for the given $r$. The survival of cooperators is ensured by approaching a growing $O$ island (such a situation is shown in the center of Fig.~\ref{snapsh2}) that is occupied quickly by the offspring of the lucky cooperator.  The resultant cooperator island is attacked simultaneously by defectors whose success is enhanced by a guerilla-type warfare fragmenting the cooperator's territory into a cluster of small shrinking islands as demonstrated in the same snapshot. This process is repeated forever if the defectors take pool-punishers off cooperator's bands with a sufficiently high probability.

The above mentioned evolutionary process implies the risk of extinction for the strategies that occur with only a low frequency. Within the region $\beta_{c4} < \beta < \beta_{c5}$ the fixation of pool-punisher will occur if the cooperators die out first during the stochastic extinction process. In the opposite case, if pool-punishers go extinct first, the cooperators have no chance to survive. Consequently, within this parameter region the system can evolve towards one of the homogeneous states where only defectors or pool-punishers are present. Henceforth, the corresponding behavior (phase) will be denoted by F, referring to fixation.

During the transient process the strategy frequencies oscillate with a growing amplitude and oscillation period, and simultaneously, the trajectory spirals out on the simplex \cite{hofbauer_98}. Similar phenomena were already reported for many other systems (for references see \cite{hofbauer_98, nowak_06, szabo_pr07}).

Figure~\ref{csr35c4} shows that the F phase is surrounded by phases exhibiting opposite behaviors when approaching the region of fixation. When approaching from the left hand side the system tends towards the O phase while from the opposite side the emergence of the D phase is favored.

\begin{figure}
\centerline{\epsfig{file=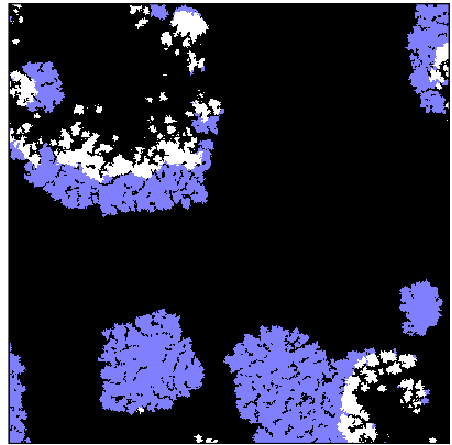,width=6cm}}
\caption{(Color online) Typical distribution of strategies within the (D+C+DO)$_c$ phase on a $400 \times 400$ portion of a larger square lattice ($L=2000$) for $r=3.5$, $\beta=0.2$ and $\gamma=0.5$. The color code is the same as used in Fig.~\ref{snapsh2}.}
\label{snapsh3}
\end{figure}

Another novel feature is the appearance of an additional three-strategy phase within a narrow range of $\beta$ values (namely, $\beta_{c2} < \beta < \beta_{c3}$ where $\beta_{c2}(r=0.4)=0.607(1)$ and $\beta_{c3}(r=0.4)=0.660(1)$). This phase will be denoted as (D+C+DO)$_c$ because the corresponding snapshot (see Fig.~\ref{snapsh3}) illustrates clearly that here the cyclic invasions occur between the D, the C, and the DO phase. This spatiotemporal structure can be reproduced very rarely because of the fast extinction of cooperators if the system is started from a random initial state even for $L>5000$. In such a case the system evolves into the DO phase that is, however, unstable against the invasion of a cooperator block with a sufficiently large size (\textit{e.g.} $10 \times 10$). It is worth mentioning that the resultant (D+C+DO)$_c$ phase will appear only after a long transient process.

The cyclic dominance of alliances has already been observed in spatial ecological models \cite{szabo_jtb07}. The present model, however, offers an interesting new example when one strategy (one species) fights continuously against a group of strategies (species), resulting in a stable stationary solution. Notice, furthermore, that the effective invasion rates between the three phases are strongly influenced by the composition and the spatiotemporal structure of the DO phase. This is one of the reasons why the fine-dependence of strategy frequencies deviates from the standard behavior discussed for the (D+C+O)$_c$ phase. The other reason is related to the effect of the pattern topology itself.

The MC simulations have confirmed clearly that the transition (at $\beta=\beta_{c2}$) from DO to (D+C+DO)$_c$ is continuous while the subsequent transition (at $\beta=\beta_{c3}$) is a weakly first-order one. The latter behavior might be related to the different time scales (characterizing the average formation and lifetimes of the competing phases) that depended on $\beta$ and $\gamma$ as well as on the synergy factor $r$.

\begin{figure}
\centerline{\epsfig{file=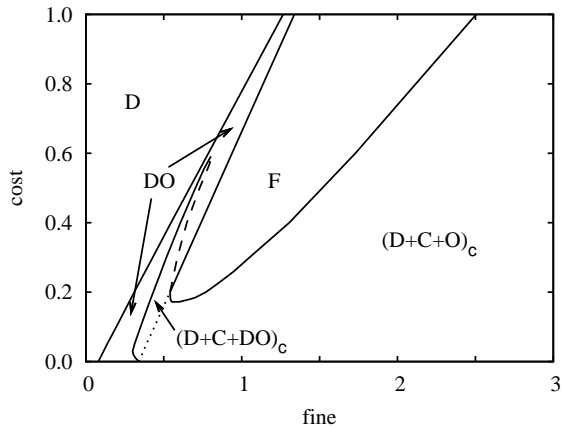,width=7.5cm}}
\caption{Full fine-cost phase diagram for $r=3.5$ and $K=0.5$. Solid (dashed) lines indicate continuous (discontinuous) phase transitions. The dotted line represents the analytic continuation of the phase boundary separating the pure D and O phases in the absence of cooperators ($C$).}
\label{phd_r35}
\end{figure}

As for the $r=2.0$ case, for $r=3.5$ the effects of different values of the punishment cost $\gamma$ on the stationary states are also studied systematically by means of MC simulations, and the results are summarized in the full fine-cost phase diagram presented in Fig.~\ref{phd_r35}.
As in the previous phase diagram (Fig.~\ref{phd_r2}), here the dotted line [separating the territory of the (D+C+DO)$_c$ and (D+C+O)$_c$ phases] indicates the transition from phase DO to O in the absence of cooperators. Notice that this dotted line is the analytic continuation of the phase boundary separating the territory of the DO phase and the fixation phase. In fact, the transition from (D+C+DO)$_c$ to (D+C+O)$_c$ is made smoother by the absence of long thermalization (between the domains of D and DO) that is due to the cyclic dominance emerging in the presence of cooperators. As a result, the transition can not be clearly identified by exclusively considering the frequencies of strategies and/or nearest-neighbor pair correlations. The difference between these phases, however, is well recognizable visually in the snapshots (compare Figs.~\ref{snapsh2} and \ref{snapsh3}). The same arguments are valid in the case of $r=2.0$ above, where the (D+C+DO)$_c$ phase was also mentioned (see the bottom panel of Fig.~\ref{phd_r2}).

Despite the many striking differences, there also exist some qualitative similarities between the fine-cost phase diagrams obtained for $r=2$ and $3.5$. Namely, for both cases only defectors remain alive if the cost $\gamma$ exceeds a threshold value (which in both cases increases fairly linearly with the increasing of the fine $\beta$). Furthermore, the system evolves into the (D+C+DO)$_c$ phase for sufficiently high fines if the cost is less than another fine-dependent threshold value. Upon decreasing of the fine a smooth transition from (D+C+O)$_c$ to (D+C+DO)$_c$ occurs when the coexistence of the $D$ and $O$ strategies is favored in the corresponding two-strategy (sub)system. The comparison of the two phase diagrams (Figs.~\ref{phd_r2} and \ref{phd_r35}) illustrates how the (D+C+DO)$_c$ phase expands together with the DO phase when $r$ increases.

The most relevant difference between the two phase diagrams is represented by the fixation allowing the formation of either the pure O or pure D phase for $r=3.5$, while only the pure O phase can occur for $r=2.0$. The mentioned difference implies significant deviation in the variation of strategy frequencies if the cost is increased for a sufficiently large fine. Namely, $\rho_C$ and $\rho_O$ vanish continuously when approaching the boundary of fixation for $r=3.5$, while the (D+C+O)$_c$ phase transform into the O phase via a first-order (discontinuous) transition if $r=2.0$. The corresponding territories (F and O phases on the fine-cost parameter plane) separate the D and (D+C+O)$_c$ phases.

\subsection{Results for the synergy factor $r=3.8$}
\label{results38}

In contrast with the values of the synergy factor $r$ considered above, here the magnitude is large-enough so that even in the absence of pool-punishers pure cooperators can survive in the presence of defectors. Accordingly, the enhancement of $r$ reduces the temptation to choose defection in the PGG, which ultimately yields a continuous increase in the cooperator frequency for the two-strategy ($D$ and $C$) system. For the present interaction graph (the square lattice) $r > r_{th1}=3.744$. In this situation the efficiency of punishment decreases alongside with the frequency of defectors who are negatively affected by the sanctions. We therefore consider the efficiency of institutionalized punishment in the case when $\rho_D/\rho_C \simeq 2$ in the absence of pool-punishers ($\rho_O=0$). For this value of $\rho_D/\rho_C$ we can extract quantitative results giving a sufficiently accurate and general picture about the impact of pool-punishment.

\begin{figure}
\centerline{\epsfig{file=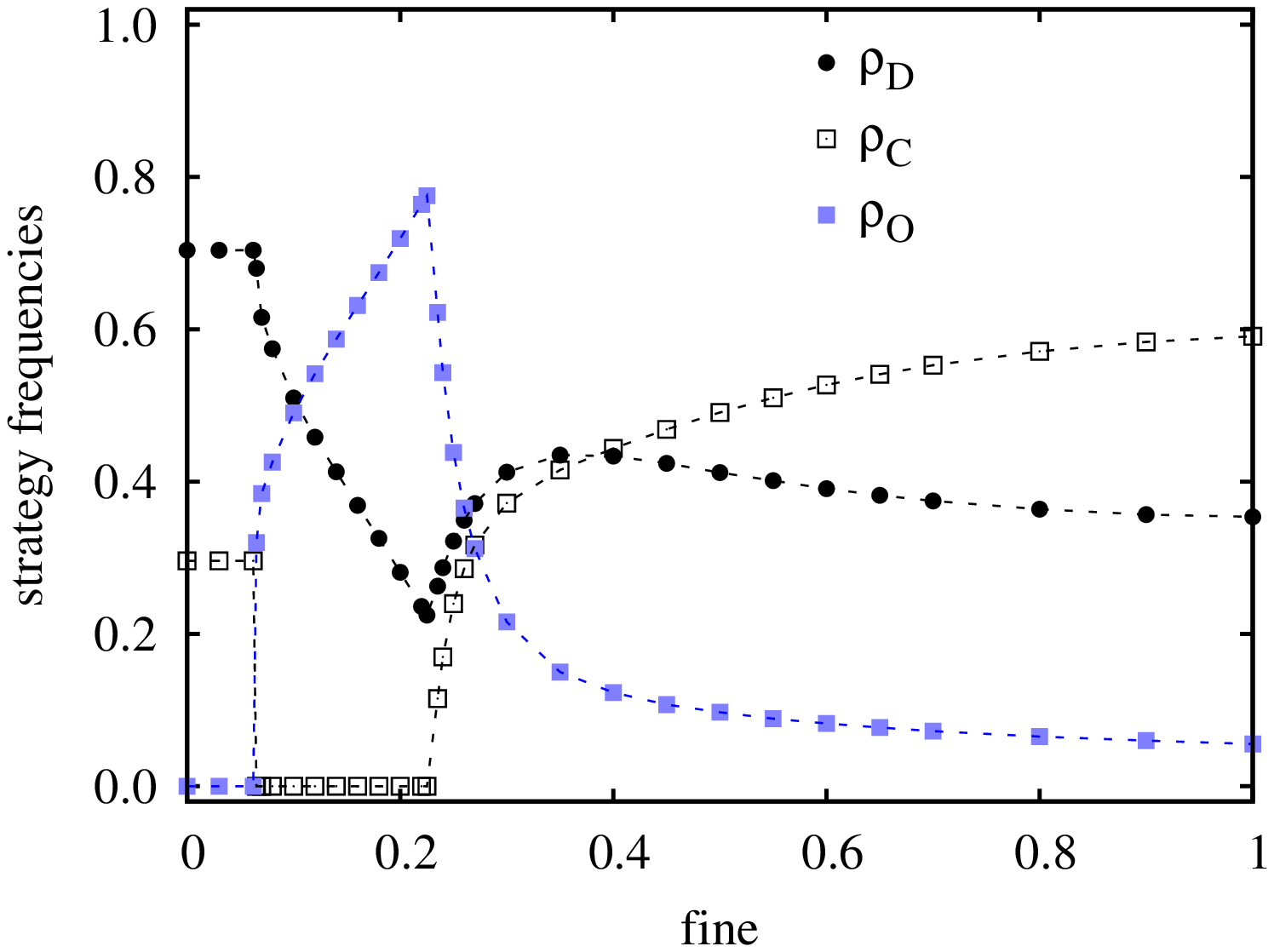,width=7.5cm}}
\centerline{\epsfig{file=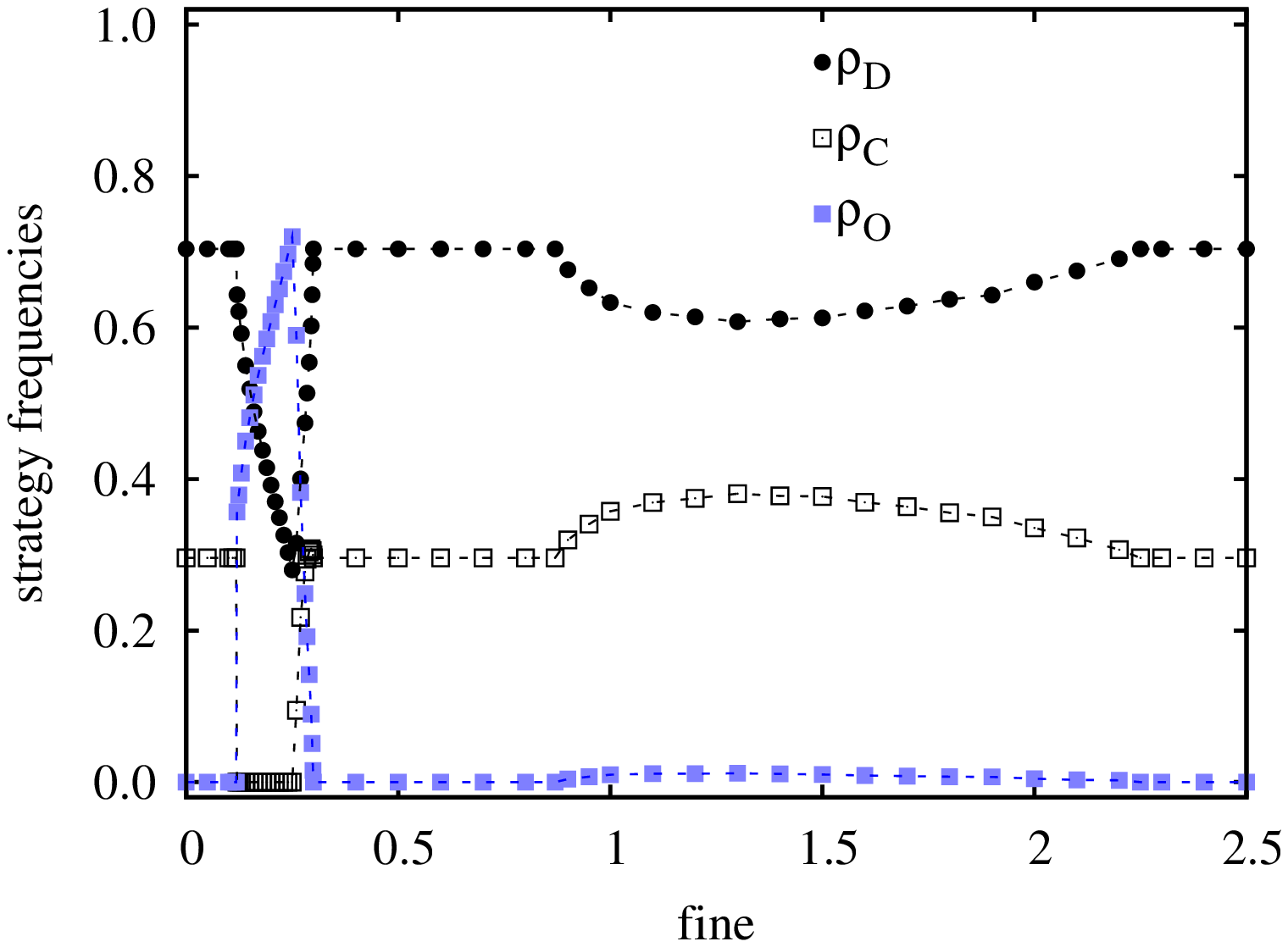,width=7.5cm}}
\caption{(Color online) Strategy frequencies {\it vs}. fine $\beta$ for the punishment cost $\gamma=0.05$ (top) and $\gamma=0.1$ (bottom) at $r=3.8$ and $K=0.5$.}
\label{cross_r38}
\end{figure}

First, we demonstrate the variations in the strategy frequencies upon increasing the fine $\beta$ for two (low) values of the cost $\gamma$ characterizing the relevant processes. Figure~\ref{cross_r38} shows that the pool punishers disappear ($\rho_O=0$) if the fine does not exceeds a threshold value increasing with the cost $\gamma$. Evidently, $\rho_D$ and $\rho_C$ remain unchanged in the absence of pool-punishers.

For both low values of the cost there exist a region of fine where cooperators die out and pool-punishers maintain cooperation at a level that increases with the fine.

The replacement of the DC phase by the DO phase happens via a first-order (discontinuous) phase transition, which is a manifestation of a more general phenomenon. Namely, the cooperative $C$ and $O$ strategies fight separately against defectors ($D$ players) and the final output depends on the success of these $D-C$ and $D-O$ struggles. Accordingly, the ``indirect territorial battle'' between $O$ and $C$ strategies will determine the final output of the game. This mechanism has already been observed for peer-punishment \cite{helbing_ploscb10} and in the spatial public goods game with reward \cite{szolnoki_epl10}.

Paradoxically, the further increase of punishment fine $\beta$ enhances the chance of cooperators to survive. Consequently, the system behavior transforms into the cyclic-dominance-governed (D+C+DO)$_c$ phase. For sufficiently high values of the fine $\beta$, however, the defectors are not capable to survive within the domains of pool-punishers, and accordingly, the (D+C+DO)$_c$ phase evolves into the (D+C+O)$_c$ phase as described above. On the contrary, for higher values of the punishment cost $\gamma$, the (D+C+DO)$_c$ phase transforms into the DC phase via a continuous transition, as illustrated in the lower panel of Fig.~\ref{cross_r38}. The MC simulations indicate that within a narrow range of cost values the (D+C+O)$_c$ phase can occur and vanish again continuously if the fine is further increased.

\begin{figure}
\centerline{\epsfig{file=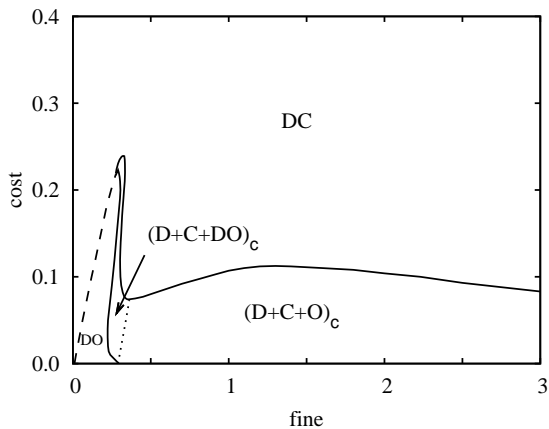,width=7.5cm}}
\caption{Full fine-cost phase diagram for $r=3.8$ and $K=0.5$. Solid (dashed) lines indicate continuous (discontinuous) phase transitions. The dotted line represents the analytic continuation of the phase boundary separating the pure D and O phases in the absence of cooperators ($C$).}
\label{phd_r3}
\end{figure}

From systematic numerical investigations for several different values of $\gamma$, we obtain the full fine-cost phase diagram that is presented in Fig.~\ref{phd_r3}. In agreement with our expectations, the pool-punishers cannot support the cooperation (and die out) if the cost of punishment exceeds a threshold value, The latter, of course, depends on the punishment fine $\beta$. In other words, at such high values of $r$ and $\gamma$ the pool-punishers become inefficient in their main task of facilitating the evolution of cooperation. This behavior is in sharp contrast with the impact of peer-punishment, which can always eliminate defectors for sufficiently high values of fine independently on the value of $r$. In case of pool-punishment the most efficient suppression of defection is achieved at the right border of the DO phase that is formed only within a limited range of fine values. As we have emphasized, for higher fines the self-organizing patterns with cyclic invasion of D, C and DO domains (the latter domains are replaced by D domains if the DO phase becomes unstable) emerge, which in case of the stable (D+C+DO)$_c$ phase manifest a new form of cyclical dominance that forms not just between individual strategies but also between strategy alliances.

\section{Summary}
\label{summary}

The impact of pool-punishment was studied in a spatial public goods games with cooperators, defectors and pool-punishers as the three competing strategies. In particular, the efficiency of pool-punishment in maintaining socially advantageous states was contrasted with that of peer-punishment \cite{helbing_njp10, helbing_ploscb10, helbing_pre10c}. For easier comparisons, in both cases the players were located on the sites of a square lattice, the collected income resulted from five five-person public good games, and the strategy evolution was governed by the same stochastic imitation rule. Monte Carlo simulations, performed for different combinations of the fine and cost of punishment at three typical values of the multiplication factor, reveal relevant differences if compared with previous results where the peer-punishers were able to dominate if the fine exceeded a threshold value that increased with the cost. Here, on the contrary, the institutional sanctions are less effective because cooperators always invade the territories of pool-punishers, even for marginally positive values of the punishment cost. On the other hand, in contrast to the well-mixed case, to maintain pool-punishment is generally viable without the necessity of sanctioning the second-order free-riders \cite{sigmund_n10}.

It turns out that the pool-punishers can dominate the system only within a strongly limited cooperator-unfriendly region of parameters. Meanwhile, for high fines the system paradoxically evolves into a self-organizing spatiotemporal pattern where the rock-paper-scissors type cyclic dominance helps the coexistence of all three strategies. In fact, we could distinguish two different cyclic phases, namely, (D+C+O)$_c$ and (D+C+DO)$_c$. The latter phase represents a new type of cyclic dominance when single strategies ($D$ and $C$) fight against an alliance ($D+O$). The possibility of these two phases governed by cyclic dominance is accompanied with an unusual sensitivity to the topological features of the self-organizing patterns. As a result, in some cases we have observed the fixation to either the homogeneous defector or the homogeneous pool-punisher phase. Based on earlier works examining the spatial PGG and its variants, the reported impact of pool punishment is expected to be robust against using different interaction graphs and group sizes.

The accurate determination of presented phase diagrams (for the large size limit) required a careful stability analysis based on the concept of competing associations \cite{szabo_pr07}. In the light of our results we can strongly recommend the application of this method for other multi-strategy systems where a complex phase diagram is likely to be encountered. A potential example is given by the present evolutionary PGG with four rather than three strategies (besides cooperators, defectors and pool-punishers containing also peer-punishers), which we wish to study in the near future.

\section*{Acknowledgments}

We thank Karl Sigmund for initiating the present investigations and stimulating discussions. This work was supported by the Hungarian National Research Fund (grant K-73449), the Bolyai Research Grant, and the Slovenian Research Agency (grant Z1-2032).


\end{document}